\documentclass[conference]{IEEEtran}

\usepackage{cite}
\usepackage{amsmath,amssymb,amsfonts}
\usepackage{graphicx}
\usepackage{booktabs}
\usepackage{array}
\usepackage{xcolor}
\usepackage{url}
\usepackage{stfloats}

\graphicspath{{./}{figures/}}

\newcommand{\PROTECT}{PROTECT-90}
\newcommand{\scType}{\texttt{sc\_type}}
\newcommand{\faultTarget}{\texttt{fault\_target}}
\newcommand{\ohm}{\Omega}

\begin{document}

\title{One-Cycle Fault Classification and Faulted-Line Identification on the PROTECT-90 Dataset:\\ An Initial Application Benchmark}

\author{%
\makebox[\textwidth][c]{%
\begin{tabular}{@{}c@{}}
Emad Abukhousa$^{1}$ \quad Abdulaziz Qwbaiban$^{2,3}$ \quad Saman Zonouz$^{1}$ \quad A. P. Sakis Meliopoulos$^{1}$\\
$^{1}$\textit{School of Electrical and Computer Engineering, Georgia Institute of Technology, Atlanta, GA, USA}\\
\{emadak, saman.zonouz, sakis.m\}@gatech.edu\\
$^{2}$\textit{Electrical Engineering Department, King Fahd University of Petroleum and Minerals, Dhahran, Saudi Arabia}\\
$^{3}$\textit{Interdisciplinary Research Center for Sustainable Energy Systems (IRC-SES), KFUPM, Dhahran 31261, Saudi Arabia}\\
\{qwbaiban\}@kfupm.edu.sa
\end{tabular}%
}%
}

\maketitle

\begin{abstract}
Open electromagnetic-transient datasets are beginning to make reproducible learning-based protection studies possible, but the practical use of these datasets still requires application-level benchmarks that define timing, sensing, and validation assumptions. This paper presents an initial application benchmark on the recently released \PROTECT{} dataset for two protection-oriented tasks: fault-type classification and discrete faulted-line identification. A compact one-dimensional convolutional neural network (CNN) is evaluated using post-inception windows of 0.25, 0.5, 1, and 2 cycles under strict episode-wise splitting. A non-convolutional multilayer perceptron (MLP) is also trained as an architecture-control baseline. The results show that both tasks are nearly saturated under full observability, with one-cycle test accuracies of $99.84\%$ for fault type and $100.00\%$ for line identification. The main performance variation appears under reduced observability: current-only inputs preserve line identification accuracy at $100.00\%$, whereas voltage-only inputs reduce line identification accuracy to $53.09\%$ with the CNN and $50.57\%$ with the MLP. This indicates that the limiting factor is measurement information rather than neural architecture. Additional stratified checks show stable performance across topology states and fault-resistance bins, while CPU inference contributes only $0.528$~ms to the one-cycle total decision time of $20.53$~ms.
\end{abstract}


\section{Introduction}
\label{sec:intro}

High-speed fault classification and selective line identification are core requirements for protection-oriented situational awareness. Modern digital relays and fault recorders provide synchronized voltage and current waveforms at sub-cycle resolution, which creates an opportunity for learning-based classifiers to operate directly on transient measurements. However, reported performance is often difficult to compare across studies because datasets, preprocessing, decision horizons, and sensing assumptions differ substantially. This issue is especially important for protection applications, where a model must be evaluated not only by test-set accuracy but also by the measurement window it requires before a decision can be formed. The reproducibility problem is structural: a recent scoping review found that only a small fraction of machine-learning protection studies release datasets or rely on public data, making it difficult to separate true methodological progress from dataset-specific assumptions~\cite{oelhaf2025scoping}.

The recently released \PROTECT{} dataset addresses part of this reproducibility problem. The dataset record was released through Zenodo on July 1, 2026~\cite{kordowich2026protect90data}, and the companion dataset paper documents the topology, waveform structure, metadata, and intended benchmark use~\cite{oelhaf2026protect90}. The present work provides an initial post-release benchmark on \PROTECT{} for early-window fault classification and discrete faulted-line identification under controlled measurement-observability assumptions, complementing recent companion studies that address continuous fault localization and broader measurement degradation on the same dataset.

Several closely related studies already establish the importance of this dataset family. Oelhaf \emph{et al.} evaluated machine-learning methods for fault detection and faulted-line identification in a double-line EMT benchmark at ICASSP 2025~\cite{oelhaf2025icassp}, making it the closest predecessor to the present work. However, that study appeared before the public release of \PROTECT{} and focused on model evaluation rather than one-cycle channel-observability ablations on the released benchmark. Later companion work benchmarked multiple models for fault classification and continuous fault localization using 10--50~ms decision windows on \PROTECT{}, showing that fault classification saturates quickly while continuous localization retains a persistent error floor~\cite{oelhaf2025controlled}. A robustness study further examined fault classification and continuous localization under missing sensors, reduced sampling, relay and substation outages, phase loss, and communication dropouts~\cite{oelhaf2026robustness}. These works are important baselines, but they leave a narrower protection question open: for the public \PROTECT{} dataset, how much measurement observability is needed for one-cycle, discrete faulted-line identification, and is any observed limitation caused by the classifier architecture or by the information available in the measured channels?

This paper addresses that question through a compact, protection-oriented application benchmark. The study is also connected to the authors' prior work on streaming-aware and latency-aware learning for power-system protection, where high offline accuracy was shown not to guarantee reliable streaming coverage or protection-grade timing~\cite{abukhousa2025wisdom,abukhousa2026latency}. A later study compared attention-based diagnostic classification with Dynamic State Estimation-Based Protection (DSE-EBP), emphasizing the distinction between fast physics-based detection and richer data-driven diagnosis~\cite{abukhousa2026transformer}. The present paper transfers that timing-conscious evaluation philosophy to an open public dataset, but narrows the scope to fault-only \PROTECT{} events and to two tasks: fault-type classification and discrete line identification.

The contributions are as follows. First, a leakage-free early-window benchmark is defined for \PROTECT{} using episode-wise splitting and post-inception windows from 0.25 to 2 cycles. Second, two task formulations are evaluated: four-class fault-type classification and four-class faulted-line identification. Third, a structured observability study compares full, current-only, voltage-only, single-location, and random channel subsets. Fourth, a flattened multilayer perceptron (MLP) baseline tests whether the observability trend is specific to the convolutional neural network (CNN) architecture. Finally, secondary checks quantify topology-conditioned performance, fault-resistance-conditioned performance, and CPU decision time.

\section{Dataset and Task Formulation}
\label{sec:dataset}

\PROTECT{} is based on a three-bus, 90~kV double-line topology with two parallel corridors: Line~1--2A/B and Line~2--3A/B~\cite{oelhaf2026protect90}. Each episode contains one short-circuit event and one second of synchronized EMT waveforms sampled at 6.4~kHz, corresponding to 128 samples per 50~Hz cycle. Three-phase voltages and currents are recorded at both terminals of each line section, giving eight relay locations and 48 waveform channels. The metadata include the fields used here as labels: \scType{} for fault type and \faultTarget{} for the faulted line segment. Because \PROTECT{} contains only fault episodes and no normal class, this paper does not study fault detection.

The two supervised tasks are defined as
\begin{align}
    y_{\mathrm{type}} &\in \{\mathrm{SLG},\mathrm{LL},\mathrm{LLG},\mathrm{LLL}\}, \\
    y_{\mathrm{line}} &\in \{1\text{-}2A,1\text{-}2B,2\text{-}3A,2\text{-}3B\}.
\end{align}
The first task asks which fault family occurred, while the second asks which physical line segment was faulted. This is a discrete line-identification task, not continuous fault-location regression along a line.

For a window length of $c$ cycles, the input tensor for episode $n$ is
\begin{equation}
    X_n^{(c)} \in \mathbb{R}^{C \times T_c}, \quad T_c = 128c,
\end{equation}
where $C$ is the number of selected channels. The window starts at the metadata fault inception time. Four windows are evaluated: 0.25, 0.5, 1, and 2 cycles, corresponding to 32, 64, 128, and 256 samples, or 5, 10, 20, and 40~ms. Episodes with faults shorter than the maximum 2-cycle window are excluded from the main benchmark so that all window lengths are comparable on the same active-fault subset. The final held-out test set contains 1,230 episodes. Splitting is performed at the episode level using a 70/15/15 train/validation/test partition stratified jointly by fault type and line label, preventing windows or prefixes from the same physical episode from crossing between splits.

\begin{figure}[t]
\centering
\includegraphics[width=\linewidth]{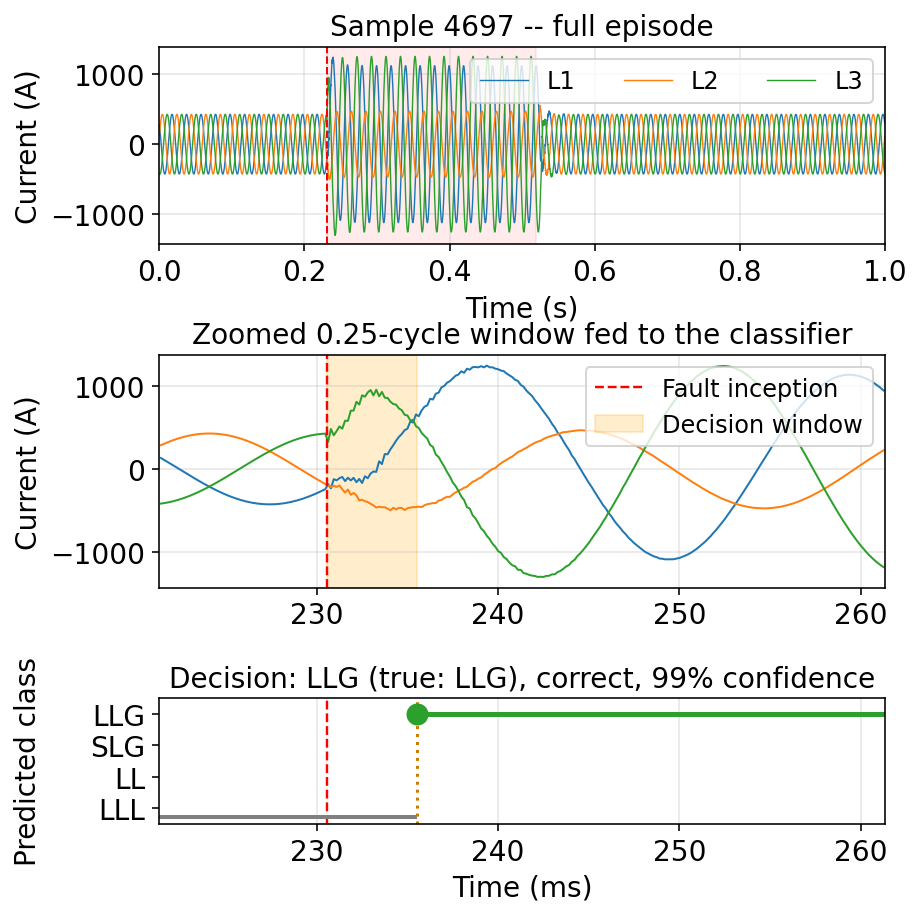}
\caption{Single-event early-decision example using a 0.25-cycle post-fault window. The top panel shows the full 1-s current waveform, the middle panel shows the zoomed waveform segment fed to the classifier, and the bottom panel shows the predicted class.}
\label{fig:single_example}
\end{figure}

Fig.~\ref{fig:single_example} illustrates the early-window decision protocol for a representative LLG event. The model is not allowed to issue the shown decision until after the selected post-inception observation window has been acquired.

\section{Benchmark Methodology}
\label{sec:method}

The primary model is a compact one-dimensional CNN operating directly on waveform windows in $(\mathrm{channels},\mathrm{time})$ format. The network consists of three temporal convolution blocks with 32, 64, and 128 filters, kernel lengths 7, 5, and 3, ReLU activations, batch normalization, two max-pooling layers, adaptive average pooling, and a final linear classifier. This architecture is intentionally small because the goal is not to introduce a new model but to define a reproducible application benchmark.

A flattened MLP baseline is used as an architecture-control model for the one-cycle observability experiments. The MLP receives the same window as the CNN after flattening and uses two hidden layers of widths 512 and 128 with ReLU activations and dropout. Because the MLP has no temporal convolution, agreement between the CNN and MLP trends indicates that the result is driven by the measured information rather than by the convolutional inductive bias.

All models are trained with cross-entropy loss, Adam optimization with learning rate $10^{-3}$, batch size 64, and early stopping on validation macro-F1. The CNN window-length benchmark allows up to 30 epochs, while the MLP baseline allows up to 25 epochs. Accuracy and macro-F1 are reported on the held-out episode test set. For the observability study, five channel configurations are evaluated: all 48 voltage/current channels, current-only 24 channels, voltage-only 24 channels, one local relay location with six channels, and one random 24-channel subset.

CPU inference latency is measured after warm-up with batch size one. The total decision time is defined as
\begin{equation}
    T_{\mathrm{decision}} = T_{\mathrm{window}} + T_{\mathrm{infer}},
\end{equation}
where $T_{\mathrm{window}}$ is the physical waveform acquisition interval and $T_{\mathrm{infer}}$ is the measured forward-pass compute time. This distinction is important because a neural model may be computationally fast but still cannot classify before the required observation window has arrived.

\section{Results}
\label{sec:results}

\subsection{Early-Window Benchmark}

\begin{figure}[t]
\centering
\includegraphics[width=\linewidth]{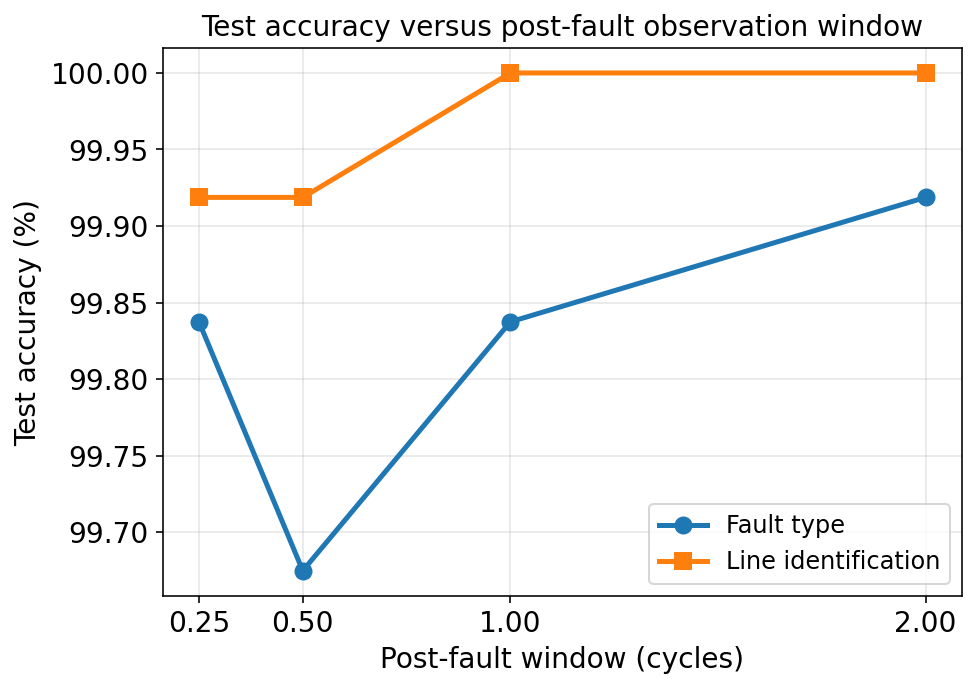}
\caption{Test accuracy versus post-fault observation window. Accuracy is already near saturated at sub-cycle windows for both tasks under full 48-channel observability.}
\label{fig:window_accuracy}
\end{figure}

Table~\ref{tab:window} summarizes the CNN benchmark across post-fault window lengths. Both tasks are nearly saturated even at 0.25 cycle. Fault-type classification reaches $99.84\%$ accuracy with a 0.25-cycle window and $99.84\%$ with a one-cycle window. Line identification reaches $99.92\%$ accuracy at 0.25 cycle and $100.00\%$ accuracy at one cycle. Fig.~\ref{fig:window_accuracy} visualizes this trend across all four window lengths for both tasks, showing that accuracy is already close to its ceiling at the shortest evaluated window. The result should therefore be interpreted as evidence that the clean \PROTECT{} tasks are highly separable under full observability, not as evidence that longer windows produce a large performance gain.

\begin{table}[t]
\centering
\caption{Early-window CNN benchmark on the held-out test set.}
\label{tab:window}
\begin{tabular}{lccc}
\toprule
Task & Window & Accuracy & Macro-F1 \\
\midrule
Fault type & 0.25 cyc. & $99.84\%$ & $99.84\%$ \\
Fault type & 0.5 cyc.  & $99.67\%$ & $99.67\%$ \\
Fault type & 1 cyc.    & $99.84\%$ & $99.84\%$ \\
Fault type & 2 cyc.    & $99.92\%$ & $99.92\%$ \\
\midrule
Line ID & 0.25 cyc. & $99.92\%$ & $99.90\%$ \\
Line ID & 0.5 cyc.  & $99.92\%$ & $99.90\%$ \\
Line ID & 1 cyc.    & $100.00\%$ & $100.00\%$ \\
Line ID & 2 cyc.    & $100.00\%$ & $100.00\%$ \\
\bottomrule
\end{tabular}
\end{table}

\begin{figure*}[t]
\centering
\includegraphics[width=\textwidth]{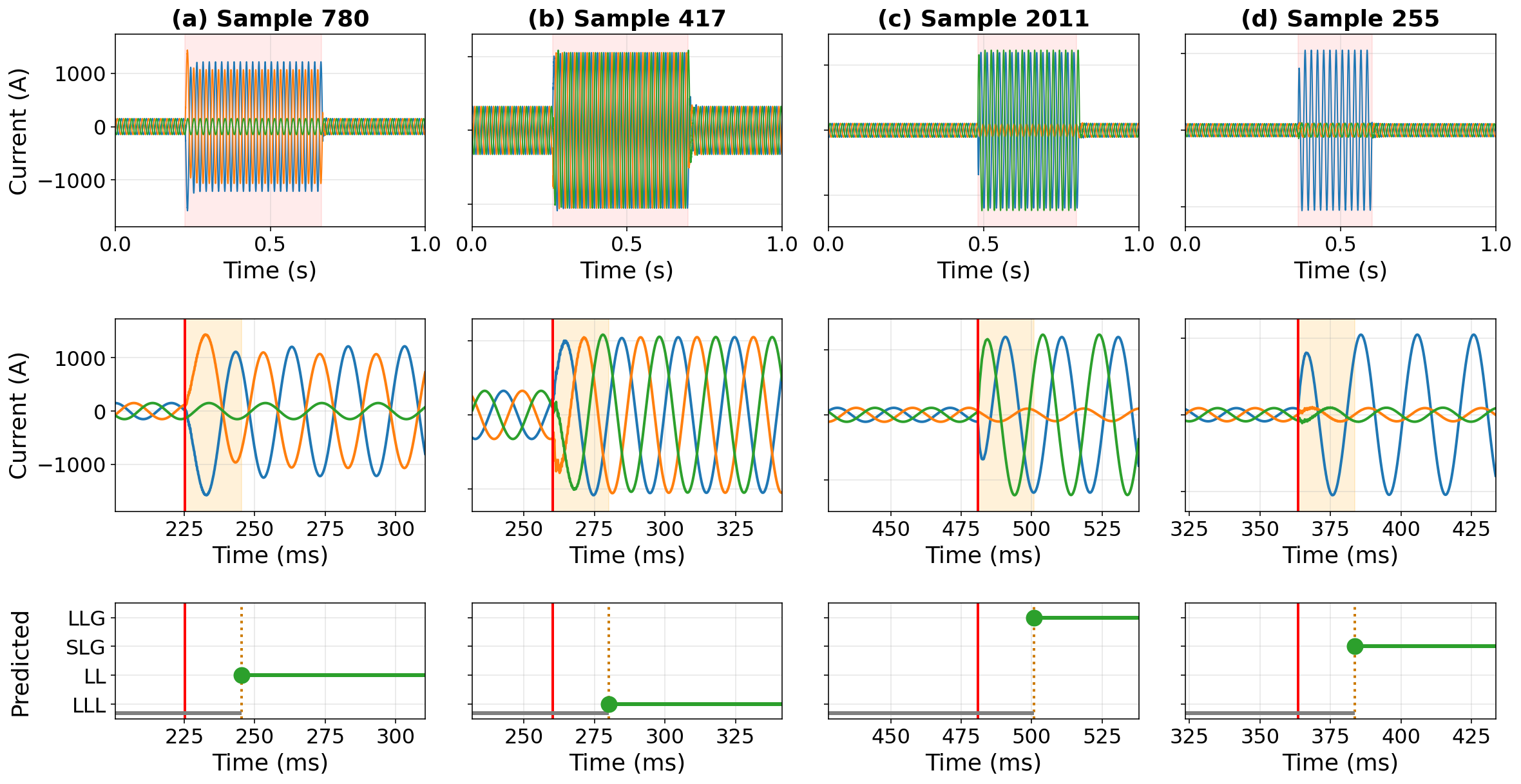}
\caption{Representative one-cycle classification examples from \PROTECT{}. Each column corresponds to a different fault event. The top row shows the full 1-s current waveform, the middle row shows the zoomed waveform segment around fault inception, and the bottom row shows the classifier output.}
\label{fig:ieee_panel}
\end{figure*}

Fig.~\ref{fig:ieee_panel} shows four representative one-cycle examples. Each column uses a different fault episode with different onset time and duration. The examples visually confirm the protocol used throughout the study: a window is collected after fault inception, then the model emits the class decision at the end of the observation interval.

\subsection{Measurement Observability}

The strongest result appears when the channel set is restricted. Table~\ref{tab:observability} shows that fault-type classification remains high under all tested channel subsets, staying above $99.10\%$ even with one local relay location. Line identification behaves differently. Current-only inputs retain $100.00\%$ accuracy, while voltage-only inputs collapse to $53.09\%$ accuracy and $37.03\%$ macro-F1. A single local relay location is also insufficient for reliable line identification, reaching only $65.12\%$ accuracy.

\begin{table}[t]
\centering
\caption{One-cycle CNN accuracy under different channel subsets.}
\label{tab:observability}
\begin{tabular}{lccc}
\toprule
Channel subset & Ch. & Fault type & Line ID \\
\midrule
All voltage + current & 48 & $99.84\%$ & $100.00\%$ \\
Current only & 24 & $99.84\%$ & $100.00\%$ \\
Voltage only & 24 & $99.59\%$ & $53.09\%$ \\
Single relay location & 6 & $99.11\%$ & $65.12\%$ \\
Random subset & 24 & $99.59\%$ & $99.02\%$ \\
\bottomrule
\end{tabular}
\end{table}

This asymmetry is the main application insight of the paper. Fault type is primarily determined by the phase involvement pattern, which remains visible in many channel configurations. Discrete line identification in the double-line topology is more dependent on branch-specific information. Because parallel line sections share terminal buses, voltage waveforms at common buses can be similar for faults on different parallel branches, while branch currents carry line-specific direction and distribution information.

\subsection{MLP Architecture Check}

Table~\ref{tab:mlp} compares the CNN to the flattened MLP for the one-cycle all-channel, current-only, and voltage-only configurations. The MLP follows the same trend as the CNN. For line identification, all-channel and current-only inputs remain near perfect, while voltage-only inputs collapse to $50.57\%$ accuracy and $29.56\%$ macro-F1. Therefore, the voltage-only degradation is not a failure of temporal convolution. It is a measurement-information limitation.

\begin{table}[t]
\centering
\caption{CNN and MLP one-cycle test accuracy under selected channel subsets.}
\label{tab:mlp}
\begin{tabular}{llcc}
\toprule
Task & Channel subset & CNN & MLP \\
\midrule
Fault type & All 48 channels & $99.84\%$ & $99.84\%$ \\
Fault type & Current only & $99.84\%$ & $100.00\%$ \\
Fault type & Voltage only & $99.59\%$ & $99.19\%$ \\
\midrule
Line ID & All 48 channels & $100.00\%$ & $99.76\%$ \\
Line ID & Current only & $100.00\%$ & $100.00\%$ \\
Line ID & Voltage only & $53.09\%$ & $50.57\%$ \\
\bottomrule
\end{tabular}
\end{table}

\subsection{Conditioned Checks and CPU Decision Time}

The one-cycle all-channel classifier was also stratified by topology state and fault resistance. Across topology flags, fault-type accuracy remained between $99.68\%$ and $100.00\%$, and line identification remained at $100.00\%$ in all states. Across four resistance bins covering the dataset's documented $0.1$--$10~\Omega$ range, fault-type accuracy ranged from $99.35\%$ to $100.00\%$, and line identification again remained at $100.00\%$. These results are useful consistency checks, but they are not the primary source of performance variation.

CPU inference latency was measured separately from the waveform acquisition interval. The one-cycle CNN forward pass required $0.528\pm0.113$~ms on CPU. Since a one-cycle window at 50~Hz requires $20$~ms of waveform acquisition, the corresponding total decision time is $20.53$~ms. The 0.25-, 0.5-, and 2-cycle total decision times were $5.59$, $10.49$, and $40.81$~ms, respectively. Thus, for this compact model, the decision timeline is dominated by the selected observation window rather than by neural-network compute cost. Table~\ref{tab:secondary} summarizes these secondary checks.

\begin{table}[t]
\centering
\caption{Secondary one-cycle checks and CPU timing.}
\label{tab:secondary}
\begin{tabular}{lcc}
\toprule
Check & Fault type & Line ID / time \\
\midrule
Topology states & $99.68$--$100.00\%$ & $100.00\%$ \\
Fault resistance bins & $99.35$--$100.00\%$ & $100.00\%$ \\
0.25-cycle timing & $0.594$ ms infer. & $5.59$ ms total \\
1-cycle timing & $0.528$ ms infer. & $20.53$ ms total \\
2-cycle timing & $0.813$ ms infer. & $40.81$ ms total \\
\bottomrule
\end{tabular}
\end{table}

\section{Discussion}
\label{sec:discussion}

The results refine how \PROTECT{} should be viewed as a protection benchmark. Under clean full-observability conditions, both studied tasks are nearly solved with sub-cycle data, consistent with prior findings that short windows contain enough information for fault classification on this dataset family~\cite{oelhaf2025controlled}. The more important result is not the saturated aggregate accuracy, but the observability breakdown: voltage-only sensing is sufficient for fault-type classification but not for discrete line identification in the double-line topology. This finding also clarifies the relationship between this paper and related robustness studies. Prior work on \PROTECT{} showed that continuous fault localization is more sensitive than fault classification under degraded measurements~\cite{oelhaf2026robustness}; the present work studies a different target, discrete line identification, but reaches a compatible high-level conclusion, that tasks requiring spatial discrimination are much more observability-sensitive than fault-type classification. The MLP baseline strengthens this conclusion by showing that the voltage-only failure occurs even without a convolutional architecture.

The physical interpretation is aligned with classical protection reasoning. Fault-family classification depends strongly on which phases are disturbed, while line identification, especially between parallel circuits sharing terminal buses, depends on where currents flow and how they distribute through the corridor. Voltage measurements at common terminals can be electrically plausible for more than one parallel branch, while current measurements remain tied to the measured line section. The learning-based classifier therefore recovers a topology-dependent information constraint that is familiar from protection engineering rather than discovering an arbitrary dataset artifact~\cite{horowitz2023relaying}. The practical implication is that a high aggregate benchmark score is not enough to claim deployability, since a model trained on full centralized measurements can appear nearly perfect, but its behavior can change sharply when the measurement set is restricted. For the studied topology, branch-current observability is sufficient for line identification, while voltage-only observability is not. This suggests that future protection-oriented benchmarks should report not only the model family and window length but also the assumed measurement configuration, especially when the topology contains parallel circuits or common-bus voltage measurements.

The results also help define what the present benchmark is not: not a broad model-ranking study, since the CNN and MLP are used mainly to test whether the observability conclusion depends on architecture; not a continuous fault-location study, since the target is the discrete line segment; and not a robustness paper in the sense of injecting missing-data faults or additive noise during operation. Those directions are important, but the purpose here is to establish the clean early-window and observability baseline on the public dataset before adding further degradation.

Several limitations must be stated. First, \PROTECT{} is simulated and uses ideal primary voltage and current signals rather than detailed CT/VT secondary models. Second, the dataset contains only fault episodes, so detection of fault versus normal operation is outside the scope of this paper. Third, the topology is intentionally standardized and compact; conclusions should not be generalized to meshed systems without further testing. Fourth, the highest resistance bin in this study is high only relative to the dataset range of 0.1--10~$\ohm$ and should not be interpreted as a general high-impedance fault result. Finally, additive noise, sensor bias, CT saturation, GPS timing errors, and cyber-physical measurement attacks are left for future work.

Despite these limitations, the paper provides a concise public-dataset extension of the authors' prior work on streaming and latency-aware protection learning~\cite{abukhousa2025wisdom,abukhousa2026latency,abukhousa2026transformer}. The contribution is not a new neural architecture; it is a reproducible application study showing what \PROTECT{} supports, where it is already easy, and where measurement observability becomes the limiting factor.
\section{Conclusion}
\label{sec:conclusion}

This paper presented an initial application benchmark on \PROTECT{} for early fault-type classification and discrete line identification. Using episode-wise splits and post-inception waveform windows, a compact CNN achieved near-perfect performance for both tasks under full observability, including $99.84\%$ fault-type accuracy and $100.00\%$ line-identification accuracy with a one-cycle window. The central finding is that measurement observability, not architecture, topology state, or fault resistance, drives the main performance variation. Current-only measurements preserve line identification, while voltage-only measurements reduce line-identification accuracy to near chance for both CNN and MLP baselines. CPU timing further shows that one-cycle decisions are dominated by waveform acquisition time, with only about 0.53~ms of CNN inference overhead. These results establish a compact, reproducible starting point for protection-oriented \PROTECT{} studies and motivate future work on noise, measurement corruption, CT/VT effects, and broader topologies.

\bibliographystyle{IEEEtran}
\bibliography{references}

\end{document}